\documentclass[10pt,letterpaper,twocolumn,reprint]{revtex4} %% two column, final layout
\pdfoutput=1
\usepackage{amsmath}

\usepackage{graphicx}

\begin{document}

%\twocolumn[ %% activate for two-column option

\title{Room temperature Bloch surface wave polaritons}

%% For REVTeX it is possible to automate superscript and e-mail callouts with the superscriptaddress option; see REVTeX4 documentation.

\author{Giovanni Lerario,$^{1,*}$ Alessandro Cannavale,$^{1,2}$ Dario Ballarini,$^{2}$ Lorenzo Dominici,$^{1,2}$ Milena De Giorgi,$^{2}$ Marco Liscidini,$^{3}$ Dario Gerace,$^{3}$ Daniele Sanvitto,$^{2}$ Giuseppe Gigli,$^{1,2,3}$}

\address{
$^1$CBN-IIT, Istituto Italiano di Tecnologia, Via Barsanti, 73010 Lecce, Italy \\
$^2$ NNL, Istituto Nanoscienze - Cnr, Via Arnesano, 73100 Lecce, Italy \\
$^3$Dipartimento di Fisica, Universit\`a di Pavia, via Bassi 6, I-27100 Pavia, Italy \\
$^4$University of Salento, Via Arnesano, 73100 Lecce, Italy \\
$^*$Corresponding author: giovannilerario@hotmail.com
}

\begin{abstract}Polaritons are hybrid light-matter quasi-particles that have gathered a significant
attention for their capability to show room temperature and out-of-equilibrium 
Bose-Einstein condensation. More recently, a novel class of ultrafast optical devices
have been realized by using flows of polariton fluids, such as switches, interferometers 
and logical gates.
However, polariton lifetimes and propagation distance are strongly limited by photon 
losses and accessible in-plane momenta in usual microcavity samples. In this work, 
we show experimental evidence of the formation of room temperature propagating polariton 
states arising from the strong coupling between organic excitons and a Bloch surface wave. 
This result, which was only recently predicted, paves the way for the realization of polariton 
devices that could allow lossless propagation up to macroscopic distances.
\end{abstract}

\maketitle

% ] %% activate for two-column option

Mixed light-matter excitations, also called polaritons, arise from
the non-perturbative coupling of optical transitions with large oscillator
strength and electromagnetic modes propagating or confined in semiconducting
or insulating media.\cite{andreani_review,kavokin_mc} Recently,
polariton excitations in two-dimensional geometries have been attracting
considerable interest, both for fundamental studies on out-of-equilibrium
Bose-Einstein condensation and superfluidity,\cite{kasprzak06,amo09,ciuti_carusotto2013rmp}
and for the possibility to exploit their nonlinear properties for
all-optical operations.\cite{liew08prl,polariton_transistor} Elementary
excitations can be observed in materials such as squaraine, porphyrin and cyanine dyes. 
As opposed to Wannier-type exciton polaritons, their ultra-high oscillator strength
allows for the observation of polariton effects at room temperature
and for Rabi splittings that could reach energies of the order of the excitonic transistion.

So far, most of the research on polaritons, concerning both
Wannier and organic types,\cite{skolnik98,larocca_organic}
has been mainly focused on the strong coupling to the photonic
modes of high-finesse planar microcavities, where the active medium
is embedded in a cavity layer between two high-reflectivity mirrors.
However, planar microcavities offer little freedom in engineering
the propagation of polaritons, due to their small accessible momenta.
Recently, it has been suggested that similar phenomena could
also be observed at the interface between a single Bragg mirror (DBR) 
and a homogeneous medium by exploiting strong light-matter coupling to
Bloch surface waves.\cite{yariv_book,descrovi2008} Such Bloch surface wave polaritons
(BSWP), \cite{bswpol2011} i.e. mixed excitations bound to the surface
of the Bragg mirror,\textemdash which exploit the confinement of the
optical mode due to total internal reflection as weel as a photonic bandgap\textemdash  could be exploited for prospective polariton devices with high efficiency and controlled
long range propagation, or for applications requiring high surface
sensitivity such as optical sensors, \cite{liscidin09josa,descrovi2010}
with clear advantages over alternative structures where metal deposition
is required.\cite{kaliteeski,symonds09}

\begin{figure}[htb]
\centerline{\includegraphics[width=7.5cm]{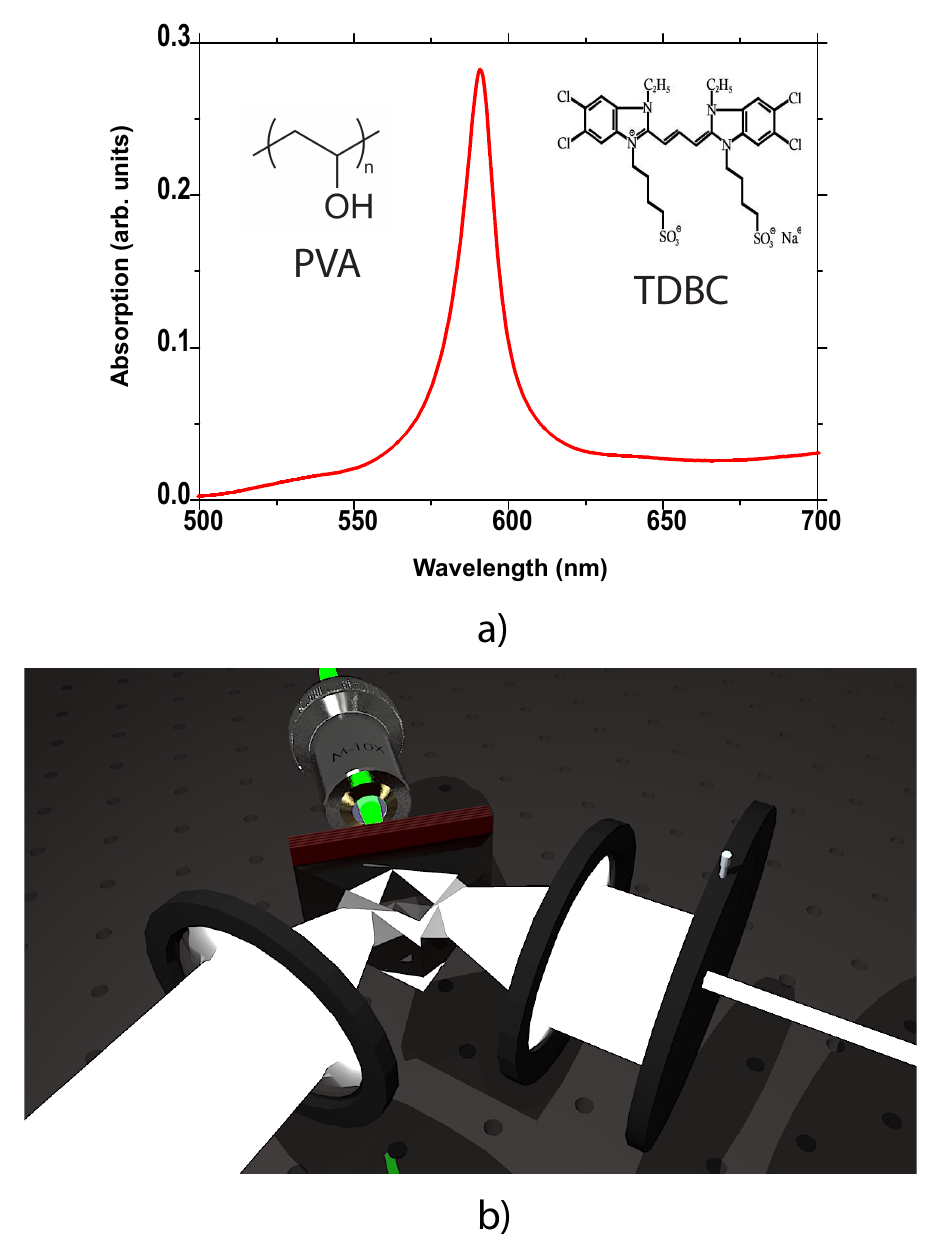}}
\caption{(a) Absorption and chemical formula of the TDBC molecule in the J-aggregate form. 
(b) Schematic drawing of the experimental setup used for reflectivity and emission measurements.}
\end{figure}

\begin{figure*}[htb]
\centerline{\includegraphics[width=13cm]{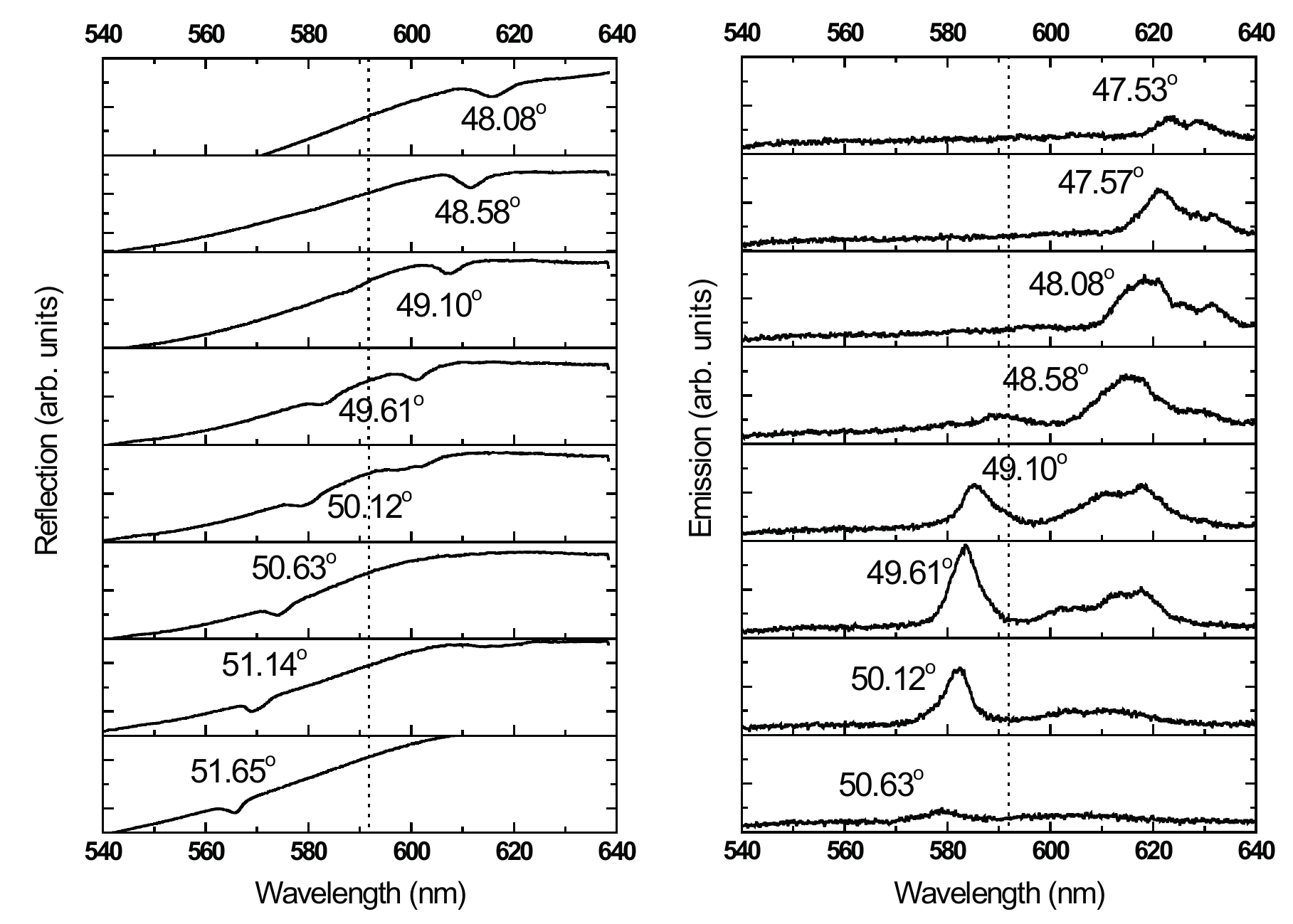}}
\caption{Reflectivity (left) and 
photoluminescence (right) spectra taken at different angles of incidence with respect to the normal to the sample inside prism.}
\end{figure*}

In this letter we report on the experimental demonstration of strong
light-matter coupling between J-aggregate excitons with Bloch
surface waves (BSW) at the interface between a SiO$_{2}$/TiO$_{2}$ Bragg
mirror and an index matching homogeneous medium. Evidence of polariton
formation is reported by independently measuring reflectivity
and emission spectra, which both show the typical anticrossing 
behavior as a function of the angle of incidence.

%\begin{figure}
%\includegraphics[scale=0.8]{Fig1.pdf}
%\protect\caption{(a) Absorption and chemical formula of the TDBC molecule in the J-aggregate form. 
%(b) Schematic drawing of the experimental setup used for reflectivity and emission measurements.}
%\end{figure}

The sample is made of five pairs of $\lambda$/4 TiO$_{2}$/SiO$_{2}$ (101 nm / 118 nm)
alternating layers deposited on a glass substrate ending with an additional
couple of TiO$_{2}$/SiO$_{2}$ (14 nm / 16 nm) layers which allow for an optimal
tuning of the BSW energy position as well as its field enhancement at
the surface of the DBR. On top of the mirror, we have spin coated
a layer of polyvinyl alcohol (PVA) (Sigma Aldrich) and 
5,6-Dichloro-2-[[5,6-dichloro-1-ethyl-3-(4-sulfobutyl)-benzimidazol-2-ylidene]-propenyl]-1-ethyl-3-(4-sulfobutyl)-benzimidazolium hydroxide, inner salt, sodium salt (TDBC) (FEW Chemicals).
The latter is a thoroughly studied cyanine dye which shows marked
property of J-aggregation both in solution and in solid state.\cite{bulovic2006}
Such characteristic results in a narrow absorption peak around 590
nm with a relative thin bandwidth compared to its monomer absorption.
The PVA matrix is a polymer with a large energy bandgap (about 5.3
eV), which is not affecting the optical properties of the TDBC. The
deposition of this last layer is obtained starting from a solution
of PVA:TDBC (1:1 weight ratio) dissolved in a mixture of methanol:water
(1:1 volume ratio). The concentration of both materials in the final
solution is of 0.5 mg/ml and additional stirring, ultrasonic bath
and heating treatments are adopted in solution. The final layer thickness is about 20 nm; we spin coated an ultrathin layer in order to retain good visibility of the modes of the system, which could be reduced due to the uncoupled molecules absorption.

Figure 1a shows the molecular structure of PVA and TDBC and the absorption
spectrum of the PVA:TDBC composite layer deposited on a glass substrate.
From the absorption spectrum is deducible the complete formation of
J-aggregates, with a full width at half maximum (FWHM) of about 45 meV and
the absence of any peaks ascribed to the TDBC monomers (usually positioned
at 520 nm). Reflectivity measurements have been done using a goniometric
system with a white ligth source focused on the sample surface and
collected at the output via two high numerical aperture lenses. Detection
of different reflection angles is done by selection of the light at
the Fourier plane as shown by the sketch in Figure 1b. The light is
finally focalized on a monochromator and the spectra are revealed
by a charged coupled device (CCD) camera. This setup allows us also
to obtain an imaging of the sample surface, so that a specific area
of interest can be selected. The overall angular resolution is about 0.2$^{\circ}$. 
The measurements are obtained in Kretschmann
configuration by means of an index matching oil. Photoluminescence, instead,
is carried out by a 532 nm cw laser exciting the PVA:TDBC from the
top side of the sample (see Figure 1b). 

\begin{figure*}[htb]
\centerline{\includegraphics[width=16cm]{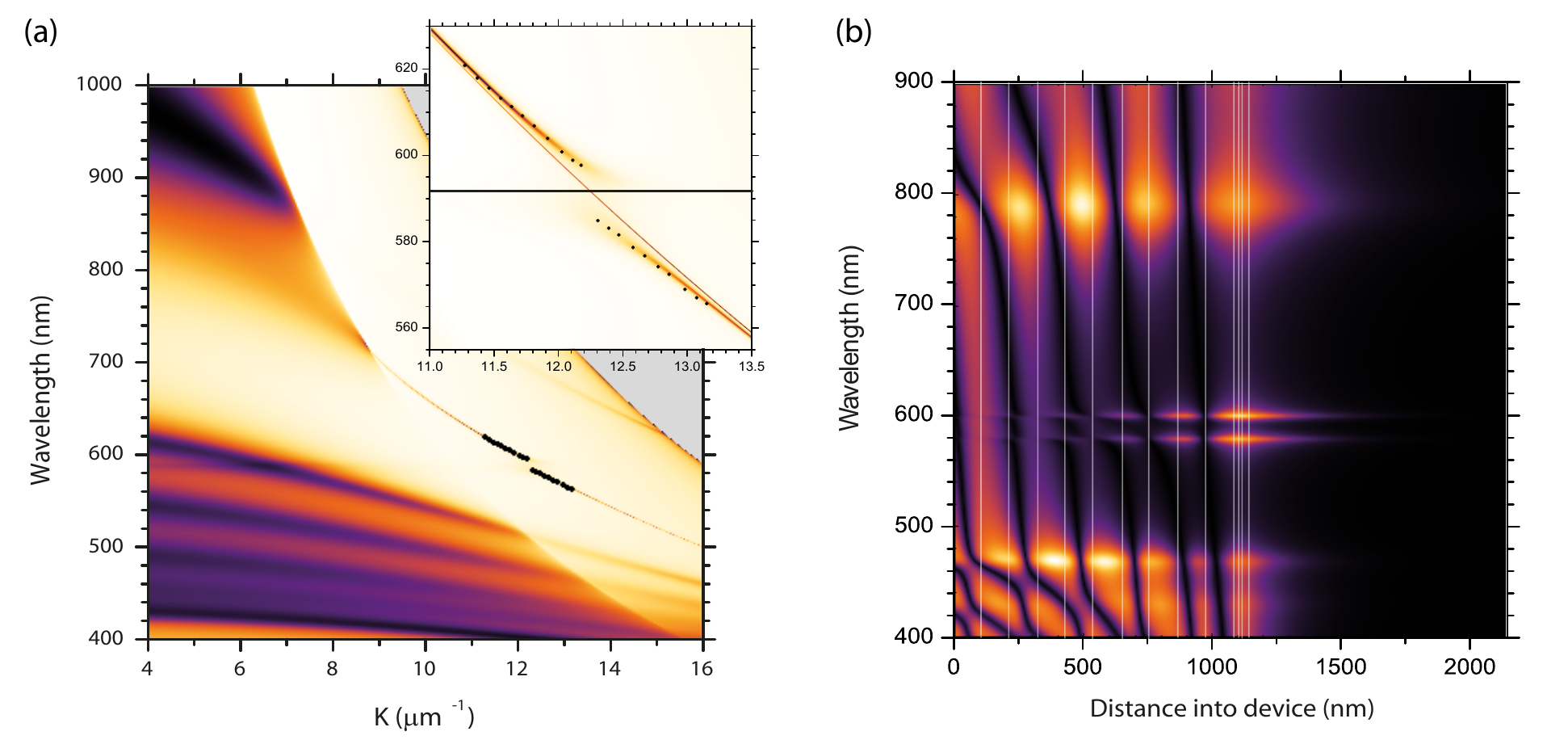}}
\caption{(a) Experimental data and theoretical calculation of the reflectivity as a function of wavelength and angle of incidence. The light line correspondes to an angle of about 42 degrees, here the Bloch surface mode starts to appear from the edge of the Bragg mirror. At k=12-12.5 $\mu m^{-1}$ the anti-crossing with the exciton of the TDBC is visible, as evidenced in the expanded inset (horizontal line is the exciton, diagonal line is the unmodified BSW mode dispersion). (b) Calculated enhancement of the electromagnetic field at 50.54 degrees as a function of wavelength and distance from the bottom of the Bragg reflector.}
\end{figure*}

%\begin{figure}
%\includegraphics[scale=0.8]{Fig2.pdf}
%\protect\caption{Reflectivity (left) and 
%photoluminescence (right) spectra taken at different angles of incidence with respect to the normal to the sample inside prism.}
%\end{figure}

Reflection and emission spectra are shown in Figure 2 for different
angles of incidence relative to the sample normal in the prism. The vertical
dotted line indicates the absorption peak energy position of the bare
TDBC aggregates. The reflectance dip (emission peak) is due to the 
excitation of a BSW at the sample surface.
As it can be seen, by tuning the BSW across the TDBC
absorption, the mode splitting into the upper and lower polariton
branches is clearly visible in both the reflectivity as well as emission
spectra, showing the typical anticrossing signature of the strong
coupling regime with a Rabi splitting of about 45 meV.

At the anticrossing point we can estimate a polariton full width at half maximum 
(FWHM) of 14 meV, while the bare BSW linewidth is below the measurement resolution
of 10 meV. This is caused by to the very strong variation of the mode energy
with angle while the selectivity is limited by the total detection
efficiency and lens aberrations. 

Theoretical calculations of the reflectivity spectra and the field
distribution, using the transfer matrix method \cite{yariv_book}, are shown in Figure 3. 

All the DBR layers parameters have been extracted from ellipsometric
measurements on the deposited materials with only very small adjustments accounting 
for thickness variation during the deposition process. We have simulated the TDBC aggregates by a single
lorentzian oscillator which fits the experimental absorption line and refractive indexes; \cite{bulovic2006,Pascual2003}
the effective refractive index, dependent on the cross section ratio between the two materials, is then optimized
in order to match the layer thickness and the Rabi splitting deduced from Figure 2.

In Figure 3a the experimental values of the reflectivity spectra in wavelength-momentum space have been plotted
together with the simulated data of the sample reflectance\textemdash in
this case including also the region within the lightcone in air\textemdash .
As it can be seen, by increasing the momentum,
the DBR stop band (visible as a bright region) shifts towards higher
energy.

 Once the total internal
reflection angle is reached (at about 42 degrees in glass) the BSW mode starts
to appear at the edge of the mirror stop-band, showing a much more abrupt
dispersion. When the mode reaches
the energy of the TDBC absorption peak (horizontal black line in the inset) the
splitting due to the strong coupling between the BSW mode and the
exciton of the dye starts to appear in perfect accordance with the
experimental data of Figure 2. The coupling appears as a deviation from the purely optical BSW (continuous diagonal line in the inset in Figure 3a).

Within the theoretical calculations we obtain a polariton FWHM band of 14 meV, in agreement with the experimental
value, while the BSW bandwidth is only 5 meV, which suggests that the
10 meV obtained experimentally are effectively given by limitations
in the overall setup resolution.

%\begin{figure}
%\includegraphics[scale=0.85]{Fig3.pdf}
%\protect\caption{(a) Experimental data and theoretical calculation of the reflectivity as a function of wavelength and angle of incidence. The light line correspondes to an angle of about 42 degrees, here the Bloch surface mode starts to appear from the edge of the Bragg mirror. At k=12-12.5 $\mu m^{-1}$ the anti-crossing with the exciton of the TDBC is visible, as evidenced in the expanded inset (horizontal line is the exciton, diagonal line is the unmodified BSW mode dispersion). (b) Calculated enhancement of the electromagnetic field at 50.54 degrees as a function of wavelength and distance from the bottom of the Bragg reflector. }
%\end{figure}

%\begin{figure}
%\includegraphics{Fig4.pdf}
%\protect\caption{Calculated enhancement of the electromagnetic field at 50.54 degrees 
%as a function of wavelength and distance from the bottom of the Bragg reflector. 
%A 27 fold enhancement of the BSWP is visible at the boundary between the last SiO$_2$ layer and the TDBC.}
%\end{figure}

Finally, in Figure 3b we have plotted the field enhancement of the BSW as a
function of wavelength and distance from the sample surface taken
at an angle of incidence of 50.45 degrees. The vertical white 
lines indicate the order of the layers composing the DBR mirror. The final phase matching layers of the DBR concentrate
the highest intensity of the electric field of the surface mode and, as
can be seen in Figure 3b, allow for the best coupling
into the 20 nm thick PVA:TDBC final layer. 

In conclusion, we have experimentally demonstrated the room temperature
formation of Bloch surface wave polaritons at the interface between a Bragg mirror and
a homogeneous medium. Such quasi-particles arise from the strong
coupling between the optical transition of an organic J-aggregate
compound deposited on top of the Bragg mirror surface, and the Bloch
surface mode existing at the interface. Given the increasing interest
for organic polaritons, and the potentialities of surface waves for
sensing, these results will stimulate further research on strongly
coupling two-dimensional layered materials to suitably engineered
surface photonic modes and long lasting polariton waves.

%\pagebreak

%\pagebreak
%\clearpage

\end{document}